\documentclass[aps,twocolumn,floats,pre]{revtex4}
\usepackage{graphics,graphicx,epsfig}
\usepackage{amssymb}
\usepackage{epsf,epstopdf,wrapfig}

\newcommand{\vf}{\mathbf{f}}
\newcommand{\vtheta}{{\mathbf\theta}}

\newcommand{\vsigma}{{\bf\sigma}}

\newcommand{\fig}[1]{Figure~\ref{#1}}

\newcommand{\R}{\mathbb{R}}

\newcommand{\LL}{\mathrm{L}}
\newcommand{\A}{\mathcal{A}}

\begin{document}

\title{Faster solutions of the inverse pairwise Ising problem}

\author{Tamara Broderick,$^a$ Miroslav Dud\'ik,$^b$ Ga\v{s}per Tka\v{c}ik,$^c$ Robert E. Schapire$^b$ and William Bialek$^{c,d}$}

\affiliation{$^a$Department of Mathematics,
$^b$Department of Computer Science,
$^c$Joseph Henry Laboratories of Physics,
$^c$Lewis--Sigler Institute for Integrative Genomics, and
$^d$Princeton Center for Theoretical Physics,
Princeton University,
Princeton, NJ 08544}

\date{\today}

\begin{abstract}
Recent work has shown that probabilistic models based on pairwise interactions---in the simplest case, the Ising model---provide surprisingly accurate descriptions of experiments on real biological networks ranging from neurons to genes.    Finding these models  requires us to solve an inverse problem:  given experimentally measured expectation values, what are the parameters of the underlying Hamiltonian?
This problem sits at the intersection of statistical physics and machine learning, and we suggest that more efficient solutions are possible by merging ideas from the two fields.  We use a combination of recent coordinate descent algorithms with an adaptation of the histogram Monte Carlo method, and implement these techniques to take advantage of the sparseness found in data on real neurons.  The resulting algorithm learns the parameters of an Ising model describing a network of forty neurons within a few minutes.  This opens the possibility of analyzing much larger data sets now emerging, and thus testing hypotheses about the collective behaviors of these networks.\end{abstract}

\maketitle

\section{Introduction} \label{sec:intro}

In the standard problems of statistical mechanics, we begin with the definition of the Hamiltonian and proceed to calculate the expectation values or correlation functions of various observable quantities.  In the inverse problem, we are given the expectation values and try to infer the underlying Hamiltonian.   
The history of the inverse problems goes back (at least) to 1959, when Keller and Zumino \cite{keller+zumino_59} showed that, for a classical gas, the temperature dependence of the second virial coefficient determines the interaction potential between molecules uniquely, provided that this potential is monotonic.  Subsequent work on classical gases and fluids considered the connection between pair correlation functions and interaction potentials in various approximations \cite{kunkin+frisch_69}, and more rigorous constructions of Boltzmann distributions consistent with given spatial variations in density  \cite{chayes+al_84} or higher order correlation functions \cite{caglioti+al_06}.

In fact the inverse problem of statistical mechanics arises in many different contexts, with several largely independent literatures.  In computer science, there are a number of problems where we try to learn the probability distribution that describes the observed correlations among a large set of variables in terms of some (hopefully) simpler set of interactions. Many algorithms for solving these learning problems rest on simplifications or approximations that correspond quite closely to established approximation methods in statistical mechanics \cite{yedidia+al_05}.  More explicitly, in the context of neural network models \cite{hopfield_82,amit_89}, a family of models referred to as `Boltzmann machines' lean directly on the mapping of probabilistic models into statistical physics problems, identifying the parameters of the probabilistic model with the coupling constants in an Ising--like Hamiltonian \cite{hinton+sejnowski_86}.

Inverse problems in statistical mechanics have received new attention because of attempts to construct explicit network models of biological systems.  Physicists have long hoped that the collective behavior which emerges from statistical mechanics could provide a model for the emergence of function in biological systems, and this general class of ideas has been explored most fully for networks of neurons \cite{hopfield_82,amit_89}.   Recent work shows how these ideas can be linked much more directly to experiment \cite{schneidman+al_06,tkacik+al_06}  by searching for maximum entropy models that capture some limited set of measured correlations.  At a practical level, implementing this program requires us to solve a class of inverse problems for Ising models with pairwise interactions among the spins, and this is the problem that we consider here.

To be concrete, we consider a network of neurons.  Throughout the brain, neurons communicate by generating discrete, identical electrical pulses termed action potentials or spikes.  If we look in a small window of time, each neuron either generates a spike or it does not, so that there is a natural description of the instantaneous state of the network by a collection of binary or Ising variables; $\sigma_{\rm i} = +1$ indicates that neuron $\rm i$ generates a spike, and $\sigma_{\rm i} = -1$ indicates that neuron $\rm i$ is silent.  Knowing the average rate at which spikes are generated by each cell is equivalent to knowing the expectation values $\langle \sigma_{\rm i}\rangle$ for all $\rm i$.  Similarly, knowing the probabilities of coincident spiking (correlations) among all  pairs of neurons is equivalent to knowing the expectation values $\langle \sigma_{\rm i}\sigma_{\rm j}\rangle$.  Of course there are an infinite number of probability distributions $P(\vsigma )$ over the states of the whole system ($\vsigma \equiv \{ \sigma_{\rm i}\}$) that are consistent with these expectation values, but if we ask for the distribution that is as random as possible while still reproducing the data---the maximum entropy distribution---then this has the form of an Ising model with pairwise interactions:
\begin{equation}
P(\vsigma ) = {1\over Z} \exp\left[ - \sum_{\rm i} h_{\rm i} \sigma_{\rm i} - {1\over 2} \sum_{\rm i\neq j} J_{\rm ij} \sigma_{\rm i}\sigma_{\rm j}\right] .
\label{ising1}
\end{equation}
The inverse problem is to find the ``magnetic fields'' $\{ h_{\rm i}\}$ and ``exchange interactions'' $\{ J_{\rm ij}\}$ that reproduce the observed values of  $\langle \sigma_{\rm i}\rangle$ and $\langle \sigma_{\rm i}\sigma_{\rm j}\rangle$.    

The surprising result of Ref \cite{schneidman+al_06} was that this Ising model provides an accurate quantitative description of the combinatorial patterns of spiking and silence observed in groups of order $N=10$ neurons in the retina as it responds to natural sensory inputs, despite only taking account of pairwise interactions.  The Ising model allows us to understand how, in this system, weak correlations among pairs of neurons can coexist with strong collective effects at the network level, and this is even clearer as one extends the analysis to larger groups (using real data for $N=40$, and extrapolating to $N=120$), where there is a hint that the system is poised near a critical point in its dynamics \cite{tkacik+al_06}.  Since the initial results, a number of groups have found that the maximum entropy models provide surprisingly accurate descriptions of other neural systems \cite{shlens+al_06,other_neurons,other_neurons_2} and similar approaches have been used to look at biochemical and genetic networks \cite{lezon+al_06,tkacik_07}.

The promise of the maximum entropy approach to biological networks is that it builds a bridge from easily observable correlations among pairs of elements to a global view of the collective behavior that can emerge from the network as a whole.  Clearly this potential is greatest in the context of large networks.  Indeed, even for the retina, methods are emerging that make it possible to record simultaneously from hundreds of neurons \cite{litke+al_04,segev+al_04}, so just keeping up with the data  will require methods to deal with much larger instances of the inverse problem.  The essential difficulty, of course, is that once we have a large network, even checking that a given set of parameters $\{h_{\rm i} , J_{\rm ij}\}$ reproduce the observed expectation values requires a difficult calculation.  In Ref \cite{tkacik+al_06} we took an essentially brute force Monte Carlo approach to this part of the problem, and then adjusted the parameters to improve the match between observed and predicted expectation values using a relatively naive algorithm.   

In this work we combine several ideas---taken both from statistical physics and from machine learning \cite{ML}---which seem likely to help arrive at more efficient solutions of the inverse problem for the pairwise Ising model.  First, we adapt the histogram Monte Carlo method  \cite{ferrenberg+swendsen_88} to `recycle' the Monte Carlo samples that we generate as we make small changes in the parameters of the Hamiltonian.  
Second, we use a coordinate descent method to adjust the parameters \cite{dudik+al_04}. 
Finally,  we exploit the fact that neurons use their binary states in  a very asymmetric fashion, so that silence is much more common that spiking.  Combining these techniques, we are able to solve the inverse problem for $N=40$ neurons in tens of minutes, rather than many days for the naive approach, holding out hope for generalization to yet larger problems.

\section{Ingredients of our algorithm}
\label{sec:ising}

\subsection{Basic formulation}

Our overall goal is to build a model for the distribution $P(\vsigma )$
over the states of the a system with $N$ elements, $\vsigma \equiv \{\sigma_1 , \sigma_2, \cdots , \sigma_N\}$.  As ingredients for determining this model, we use low order statistics computed from a set of $m$ samples
$\{ \vsigma^1 , \vsigma^2 , \cdots , \vsigma^m$\}, which we can think of as samples drawn from the distribution $P(\vsigma )$.
The classical idea of maximum entropy models  is that we should construct $P(\vsigma )$ to generate the correct values of certain average quantities (e.g., the energy in the case of the Boltzmann distribution), but otherwise the distribution should be `as random' as possible \cite{jaynes_57}.  Formally this means that we find $P(\vsigma )$ as the solution of a constrained optimization problem, maximizing the entropy of the distribution subject to conditions that enforce the correct expectation values.  We will refer to the quantities whose averages are constrained as ``features'' of the system, ${\bf f} \equiv \{ f_1 , f_2 , \cdots , f_K\}$, where each $f_\mu$ is a function of the state $\vsigma$, $f_\mu (\vsigma )$.  

One special set of average features are just the marginal distributions for subsets of the variables.  Thus we can construct the one--body marginals
\begin{equation}
P_{\rm i} (\sigma_{\rm i} ) = \sum_{\{\sigma_{\rm j \neq i}\}} P(\sigma _1, \sigma_2, \cdots , \sigma _N) ,
\end{equation}
the two--body marginals,
\begin{equation}
P_{\rm ij} (\sigma_{\rm i}, \sigma_{\rm j} ) = \sum_{\{\sigma_{\rm k \neq i,j}\}} P(\sigma _1, \sigma_2, \cdots , \sigma _N) ,
\end{equation}
and so on for larger subsets.  The maximum entropy distributions consistent with marginal distributions up to $K$--body terms generates a hierarchy of models that capture increasingly higher--order correlations, monotonically reducing the entropy of the model as $K$ increases, toward the true value \cite{schneidman+al_03}.

Let $\tilde P$ denote the empirical distribution
\begin{equation}
\label{eq:empirical_distr}
    \tilde P (\vsigma)=\frac1m\sum_{n=1}^m\delta(\vsigma,\vsigma^n) ,
\end{equation}
where $\delta(\vsigma,\vsigma')$ is the Kronecker delta, equal to
one when $\vsigma=\vsigma'$ and equal to zero otherwise. The
maximum-entropy problem is then
\begin{equation}
\textstyle
\label{eq:maxent}
   \max_{P} S [P]
   \text{ such that }
   \langle \vf(\vsigma)\rangle_P =\langle \vf(\vsigma)\rangle_{\tilde P}
\enspace,
\end{equation}
where $S[P]$ denotes the entropy of the distribution $P$, and $\langle \cdots\rangle_P$ denotes an expectation value with respect to that distribution. Using the method of Lagrange
multipliers,  the solution to the maximum entropy
problem has the form of a Boltzmann or Gibbs distribution \cite{jaynes_57},
\begin{equation}
 Q_{\vtheta}({\mathbf \sigma})= {1\over {{\mathbf Z}({\vtheta}) }}
\exp\left[ -\sum_{\mu =1}^K \theta_\mu f_\mu (\sigma )\right] ,
\label{qtheta}
\end{equation}
where as usual the partition function ${\mathbf Z}({\vtheta})$
is the normalization constant ensuring that the distribution
$Q_{\vtheta}$ sums to one, and the
parameters $\{\theta_1 , \theta_2 , \cdots , \theta_K\}\equiv \vtheta\in\R^{K}$ correspond
to the Lagrange multipliers.
Note that the expression for
$Q_{\vtheta}$ above describes the pairwise Ising model,
Eq (\ref{ising1}), with ${\vtheta} = \{h_{\rm i} , J_{\rm ij}\}$, and the features $\mathbf f$ are  the one--spin and two--spin combinations $\{\sigma_{\rm i}, \sigma_{\rm i}\sigma_{\rm j}\}$.

Rather than thinking of our problem as that of maximizing the entropy subject to constraints on the expectation values, we can now think of our task as searching the space of Gibbs distributions, parameterized as in Eq (\ref{qtheta}), to find the values of the parameters $\vtheta$ that generate the correct  expectation values.  Importantly, because of basic thermodynamic relationships, this search can also be formulated as an optimization problem.
Specifically, we recall that expectation values in statistical mechanics can be written as derivatives of the free energy, which in turn is the logarithm of the partition function (up to factors of the temperature, which isn't relevant here).  Thus, for distribution in the form of Eq (\ref{qtheta}), we have
\begin{equation}
\langle f_\mu ({\vsigma })\rangle_{Q_{\vtheta}} = -{{\partial\ln {\mathbf Z}({\vtheta })}\over{\partial\theta_\mu}} .
\end{equation}
Enforcing that these expectation values are equal to the expectation values computed from our empirical samples means solving the equation
\begin{equation}
\langle f_\mu ({\vsigma })\rangle_{\tilde P} \equiv {1\over m}\sum_{n=1}^m
f_\mu ({\vsigma}^n )= -{{\partial\ln {\mathbf Z}({\vtheta })}\over{\partial\theta_\mu}} .
\end{equation}
But this can be written as
\begin{widetext}
\begin{eqnarray}
{1\over m}\sum_{n=1}^m f_\mu ({\vsigma}^n ) &=& 
{1\over m}\sum_{n=1}^m {\partial\over{\partial\theta_\mu}}\sum_{\nu =1}^K \theta_\nu f_\nu({\vsigma}^n) =
-{{\partial\ln {\mathbf Z}({\vtheta })}\over{\partial\theta_\mu}} \\
0&=& {\partial\over{\partial\theta_\mu}} {1\over m}\sum_{n=1}^m\left[ -\ln {\mathbf Z}({\vtheta}) - \sum_{\nu=1}^K \theta_\nu f_\nu({\vsigma}^n)\right]\\
&=& {\partial\over{\partial\theta_\mu}} {1\over m}\sum_{n=1}^m \ln Q_{\vtheta} ({\vsigma}^n) .
\end{eqnarray}
\end{widetext}
Thus we see that matching the empirical expectation values is equivalent to looking for a local extremum (which turns out to be a maximum) of the quantity 
\begin{equation}
{1\over m} \sum_{n=1}^m \ln Q_{\vtheta} ({\vsigma}^n ).
\label{logQ}
\end{equation}
But if all the samples $\{{\vsigma}^1 , {\vsigma}^2 , \cdots , {\vsigma}^n\}$ are drawn independently, then the total probability that the Gibbs distribution with parameters $\vtheta$ generates the data is $P_{\rm total} = \prod_n Q_{\vtheta}({\vsigma}^n )$, and so the quantity in Eq (\ref{logQ}) is (up to a factor of $m$), just $\ln P_{\rm total}$.  Finding the maximum entropy distribution thus is equivalent to maximizing the probability or likelihood that our model generates the observed samples,   within the class of models defined by the Gibbs distribution Eq (\ref{qtheta}).

We recall that, in information theory \cite{cover+thomas_91}, probability distributions implicitly define strategies for encoding data, and the shortest codes are achieved when our model of the distribution actually matches the distribution from which the data are drawn.  Since code lengths are related to the negative logarithm of probabilities, it is convenient to define the cost of coding or log loss $\LL_{\tilde P}({\vtheta })$ that arises when we use the model with parameters $\vtheta$ to describe data drawn from the empirical distribution $\tilde P$:
\begin{equation}
\label{eq:logloss:1}
  \LL_{\tilde P}(\vtheta)
  = -\frac1m\sum_{n=1}^m \ln Q_{\vtheta}(\vsigma^n)
  = \langle -\ln Q_{\vtheta}(\vsigma)\rangle_{\tilde P}
\enspace.
\end{equation}
Comparing with Eq (\ref{logQ}), we obtain the \emph{dual} formulation of the maximum entropy problem,
\begin{equation}
\textstyle
\label{eq:dual}
\min_{\vtheta} \LL_{\tilde P}(\vtheta) .
\label{dual}
\end{equation}

Why is the optimization problem in Eq (\ref{dual}) difficult?  In principle, the convexity properties of free energies should make the problem well behaved and tractable.  But it remains possible for 
$\LL_{\tilde P}({\vtheta})$ to have a very sensitive dependence on the parameters $\vtheta$, and this can cause practical problems, especially if, as suggested in Ref \cite{tkacik+al_06}, the systems we want to describe are poised near a critical point.  Even before encountering this problem, however, we face the difficulty that computing $\LL_{\tilde P}({\vtheta})$ or even its gradient in parameter space involves computing expectation values with respect to the distribution $Q_{\vtheta}({\vsigma})$.  Once the space of states becomes large, it is no longer possible to do this by exact enumeration.  We can try to use approximate analytical methods, or we can use Monte Carlo methods.

\subsection{Monte Carlo methods}

Our general strategy for solving the optimization problem in Eq (\ref{dual}) will be to use standard Monte Carlo simulations \cite{MC1,MC2, GemanGe84}  to generate samples from the distribution $Q_{\vtheta} ({\vsigma})$, approximate the relevant expectation values as averages over these samples, and then use the results to propose changes in the parameters $\vtheta$ so as to proceed toward a minimum of $\LL_{\tilde P}({\vtheta})$.    Implemented naively, as in Ref \cite{tkacik+al_06}, this procedure is hugely expensive, because at each new setting of the parameters we have to generate  a new set of Monte Carlo samples.  Some of this cost can be avoided using the ideas of histogram Monte Carlo \cite{ferrenberg+swendsen_88}.

We recall that if we want to compute the expectation value of some function $\Phi ({\vsigma})$  in the distribution $Q_{\vtheta '}({\vsigma})$, this can be written as
\begin{eqnarray}
\langle \Phi ({\vsigma })\rangle_{{\vtheta '}} &\equiv& \sum_{\vsigma} Q_{\vtheta '} ({\vsigma}) \Phi({\vsigma})\\
&=&\sum_{\vsigma} Q_{\vtheta} ({\vsigma})\left[ {{Q_{\vtheta '}({\vsigma})}\over{Q_{\vtheta} ({\vsigma})}} \Phi({\vsigma})\right]
\\
&=&{\Bigg\langle} {{Q_{\vtheta '}({\vsigma})}\over{Q_{\vtheta} ({\vsigma})}} \Phi({\vsigma}) {\Bigg\rangle}_{\theta}\\
&=& 
{
{\langle \Phi({\vsigma}) \exp[ - ({\vtheta '} - {\vtheta}){\bf\cdot} {\mathbf f}({\vsigma})]\rangle_{\vtheta}}
\over
{\langle \exp[ - ({\vtheta '} - {\vtheta}){\bf\cdot} {\mathbf f}({\vsigma})]\rangle_{\vtheta}}
} ,
\label{difftheta}
\end{eqnarray}
where we denote expectation values in the distribution $Q_{\vtheta}({\vsigma})$ by $\langle \cdots \rangle_{\vtheta}$, and similarly for $\vtheta '$.  We note that  Eq (\ref{difftheta}) is exact.  The essential step of histogram Monte Carlo is to use an approximation to this equation, replacing the expectation value in the distribution $Q_{\vtheta}$ by an average over a set of samples drawn from Monte Carlo simulation of this distribution.

Consider an algorithm $\cal A$ which searches for a minimum of $\LL_{\tilde P}({\vtheta})$.  As this algorithm progresses, the values of the parameters $\vtheta$ change slowly.  We will divide these changes into stages $s=1, 2, \cdots$, and within each stage we will perform $t=1, 2, \cdots , T$ iterations.  At the first iteration, we will generate, via Monte Carlo, $M$ samples of the state  $\vsigma$ drawn out of the distribution appropriate to the current value of $\vtheta = {\vtheta}(s,t=1)$.  Let us refer to averages over these samples as $\langle \cdots \rangle_{MC{\vtheta}}$, which should approximate $\langle \cdots \rangle_{\vtheta}$.  At subsequent iterations, the parameters $\vtheta$ will be adjusted (see below for details), but we keep the same Monte Carlo samples and approximate the expectation values over the distribution with parameters ${\vtheta '}={\vtheta}(s,t)$ as
\begin{equation}
\langle \Phi({\vsigma})\rangle_{\vtheta '} \approx
{
{\langle \Phi({\vsigma}) \exp[ - ({\vtheta '} - {\vtheta}){\bf\cdot} {\mathbf f}({\vsigma})]\rangle_{MC\vtheta}}
\over
{\langle \exp[ - ({\vtheta '} - {\vtheta}){\bf\cdot} {\mathbf f}({\vsigma})]\rangle_{MC\vtheta}}
} .
\end{equation}
We denote  this approximation as $\langle \cdots \rangle_{\vtheta '} \approx
\langle \cdots \rangle_{{\vtheta '} | {\vtheta}}$.  Once we reach $t=T$, we run a new Monte Carlo simulation appropriate to the current value of $\vtheta$, and the cycle begins again at stage $s=s+1$.  This strategy is summarized as pseudocode in Fig \ref{fig:alg_shell}.

\begin{figure*}[t] 
\begin{tabbing}
{\bf Input}:~~~~\= empirical observations
                     $\vsigma^1,\cdots,\vsigma^m$ \\
   \> parameters $M$ and $T$ \\
   \> access to maxent inference algorithm $\A$\\

{\bf Algorithm}: \\
~~~~~~~~\=initialize $\A$ \\
 \>$\vtheta(1,1)\leftarrow $ initial parameter vector computed by $\A$ \\
 \>for $s=1,2,\ldots$\\
 \>~~~~~~~\=generate $M$ Monte Carlo samples using $\vtheta(s,1)$\\
 \>   \>for $t=1,\ldots,T$\\
 \>   \>~~~~~~~\=run one iteration of $\A$, approximating
$\langle \cdots \rangle_{\vtheta(s,t)} = \langle\cdots\rangle_{{\vtheta}(s,t)|{\vtheta}(s,1)}$\\
 \>   \>   \>$\vtheta(s,t+1)\leftarrow $ parameter vector computed by $\A$
                   on this iteration\\
 \>   \>$\vtheta(s+1,1)\leftarrow\vtheta(s,T+1)$
\end{tabbing}
\caption[Algorithm shell]{Pseudocode for our scheme, which reuses
a Monte Carlo sample set across $T$ iterations of a 
maxent algorithm $\A$.}
\label{fig:alg_shell}
\end{figure*}

Once we chose the optimization algorithm $\A$, there are still two free parameters in our scheme:  the number of samples $M$ provided by the
Monte Carlo simulation at each stage, and the number of iterations per stage $T$. In our
experiments, we explore how choices of these parameters influence
the convergence of the resulting algorithms.

\subsection{Parameter adjustment and the use of sparsity}

At the core of our problem is an algorithm which tries to adjust the parameters $\theta$ so as to minimize $\LL_{\tilde P}(\theta )$.    In Ref \cite{tkacik+al_06} we used a simple gradient descent method.  Here we use a coordinate descent method \cite{dudik+al_04}, adapted from Ref \cite{CollinsScSi02} and tuned specifically for the maximum entropy problem.  Where gradient descent methods attempt to find a vector in parameter space along which one can achieve the maximum reduction in $\LL$, coordinate descent methods explore in sequence the individual parameters $\theta_\mu$.  

Beyond the general considerations in Refs \cite{dudik+al_04,CollinsScSi02}, coordinate descent may be especially well suited to our problem because of the   the sparsity of
the feature vectors.   
In implementing any parameter adjustment algorithm, it is useful to take account of the fact that in networks of real neurons, spikes and silences are used very asymmetrically.  It thus is useful to write the basic variables as $n_{\rm i} = (\sigma_{\rm i} +1)/2 \in \{0,1\}$.   This involves a slight redefinition of the parameters $\{h_{\rm i} , J_{\rm ij}\}$, but once this is done one finds that individual terms $\propto n_{\rm i}$ or $\propto n_{\rm i} n_{\rm j}$ are often zero, because spikes (corresponding to $n_{\rm i} =1$) are rare.
This is true not just in the experimental data, but of course also in the Monte Carlo simulations once we have parameters that approximately reproduce the overall probability of spiking vs. silence.
This sparsity can lead to substantial time savings, since all expectation values can be evaluated in time proportional to the number of non--zero elements \cite{reweighting}.

As an alternative to coordinate descent method, we also used a general purpose convex optimization algorithm known
as the limited memory variable metric (LMVM)~\cite{tao-user-ref},
which is known to outperform others such general purpose algorithms on a wide variety of problems  \cite{Malouf02}.  While LMVM has the advantage of changing multiple parameters
at a time, the cost of updating may outweigh this advantage, especially for very sparse data such as ours.

Both the coordinate descent and the LMVM schemes are initialized with parameters [in the notation of Eq (\ref{ising1})] $J_{\rm ij} =0$ and the $h_{\rm i}$ chosen to exactly reproduce the expectation values $\langle \sigma_{\rm i}\rangle$.    Our Monte Carlo method follows the discussion of the Gibbs sampler in Ref 
\cite{GemanGe84}.   All computations were on the same computing cluster \cite{fafner} used in Ref \cite{tkacik+al_06}.

\section{Experiments} \label{sec:experiments}

In this section, we evaluate the speed of convergence of our
algorithms as a function of $M$  (the number of Monte Carlo samples drawn in a sampling round) and $T$ (the number of learning steps per stage) under a fixed time budget.
We begin with synthetic data, for which we know the correct answer, and then consider the problem of constructing maximum entropy models to describe real data.

\subsection{Synthetic data}

The maximum entropy construction is a strategy for simplifying our description of a system with many interacting elements.  Separate from the algorithmic question of finding the maximum entropy model is the natural science question of whether this model provides a good description of the system we are studying, making successful predictions beyond the set of low order correlations that are used to construct the model.  To sharpen our focus on the algorithmic problem, we use as ``empirical data'' a set of samples generated by Monte Carlo simulation of an Ising model as in Eq (\ref{ising1}).  To get as close as possible to the problems we face in analyzing real data, we use the parameters of the Ising model found in Ref \cite{tkacik+al_06} in the analysis of activity in a network of $N=40$ neurons in the retina as it responds to naturalistic movies.  We note that this model has competing interactions, as in a spin glass, with multiple locally stable states; extrapolation to larger systems suggests that the parameters are close to a critical point.

\begin{figure}[b]
\begin{center}
\includegraphics[width=\linewidth]{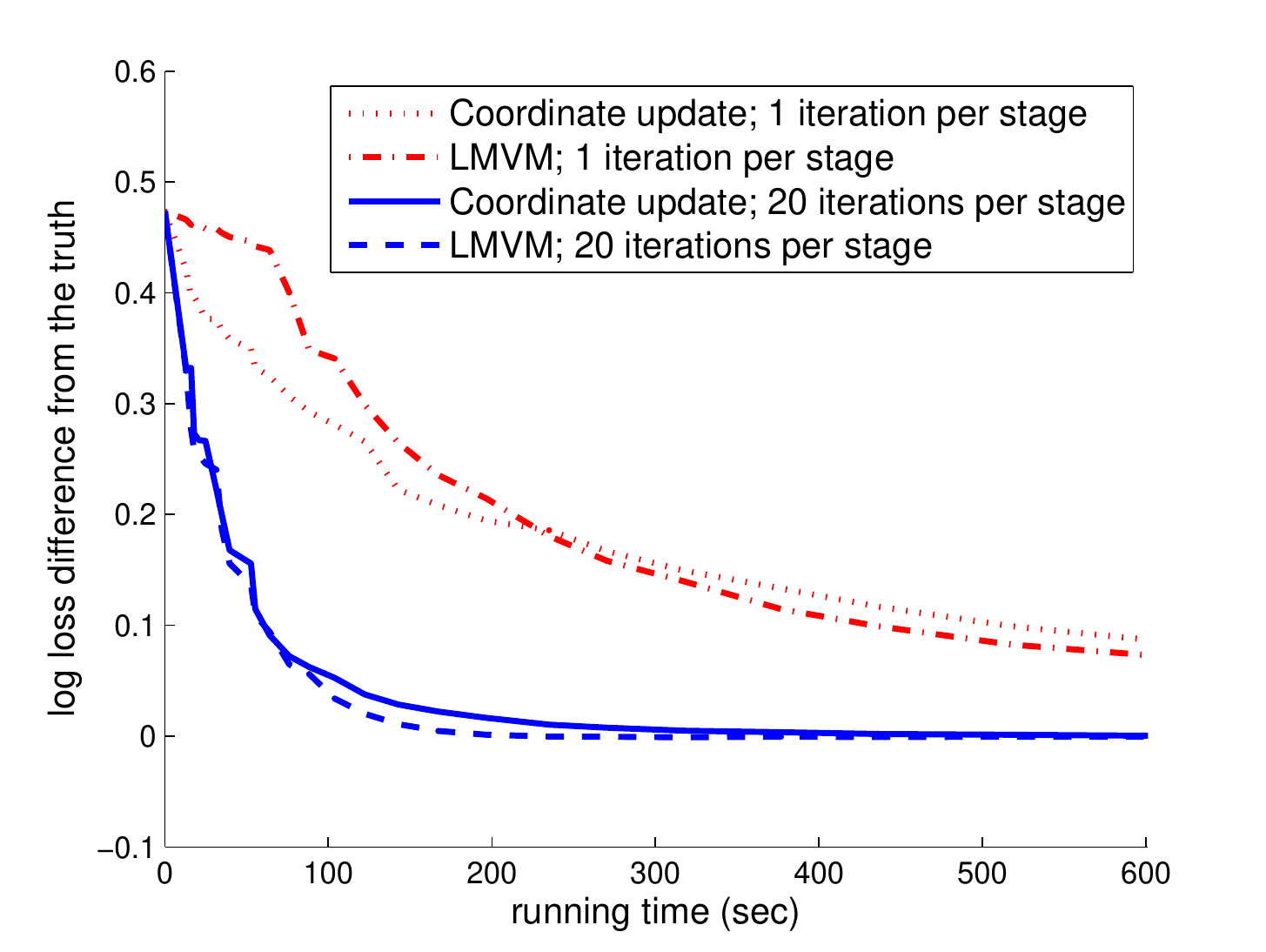}
\end{center}
\caption{\label{fig:loss_compare} 
Approximate difference in the test log loss
between the current solution of an algorithm and the exact model that generates the test data, as a function of the algorithm running time. The number of Monte Carlo samples
at each stage is set to 320,000. The two choices of optimization
algorithm exhibit similar performance, but much greater efficiency
is achieved in both cases by including multiple iterations of
learning algorithm $\A$ per stage.}
\end{figure}

Starting with the parameters ${\vtheta}_0\equiv \{h_{\rm i} , J_{\rm ij}\}$ determined in Ref \cite{tkacik+al_06}, we generate $m=2\times 10^5$ samples by Monte Carlo simulation.  Let us call the empirical distribution over these samples $\hat P$.  Then we proceed to minimize $\LL_{\hat P} ({\vtheta})$.  To monitor the progress of the algorithm, we compute 
\begin{equation}
\Delta\LL ({\vtheta}) = \LL_{\hat P}({\vtheta}) - \LL_{\hat P}({\vtheta}_0 ).
\end{equation}
Note that this computation requires us to estimate the log ratios of partition functions, for which we again use the histogram Monte Carlo approach,
\begin{eqnarray}
{{{\mathbf Z}({\vtheta})}\over{{{\mathbf Z}({\vtheta}_0)}}}
&=&
\langle \exp\left[ - ({\vtheta}-{\vtheta}_0){\bf \cdot} {\bf f}({\vsigma})\right] \rangle_{{\vtheta}_0}\\
&\approx& {\bigg\langle} \exp\left[ - ({\vtheta}-{\vtheta}_0){\bf \cdot} {\bf f}({\vsigma})\right] {\bigg\rangle}_{MC{\vtheta}_0} .
\end{eqnarray}
As a check on this approximation, we can exchange the roles of ${\vtheta}$ and ${\vtheta}_0$,
\begin{equation}
{{{\mathbf Z}({\vtheta}_0)}\over{{{\mathbf Z}({\vtheta})}}}
\approx 
 {\bigg\langle} \exp\left[ + ({\vtheta}-{\vtheta}_0){\bf \cdot} {\bf f}({\vsigma})\right] {\bigg\rangle}_{MC{\vtheta}} ,
\end{equation}
and test for consistency between the two results. Finally, to compare the performance of the two optimization algorithms, we withhold some fraction of the data, here chosen to be 10\%, for testing.

\begin{figure*}[bt]
\begin{center}
\includegraphics[width=0.45\linewidth]{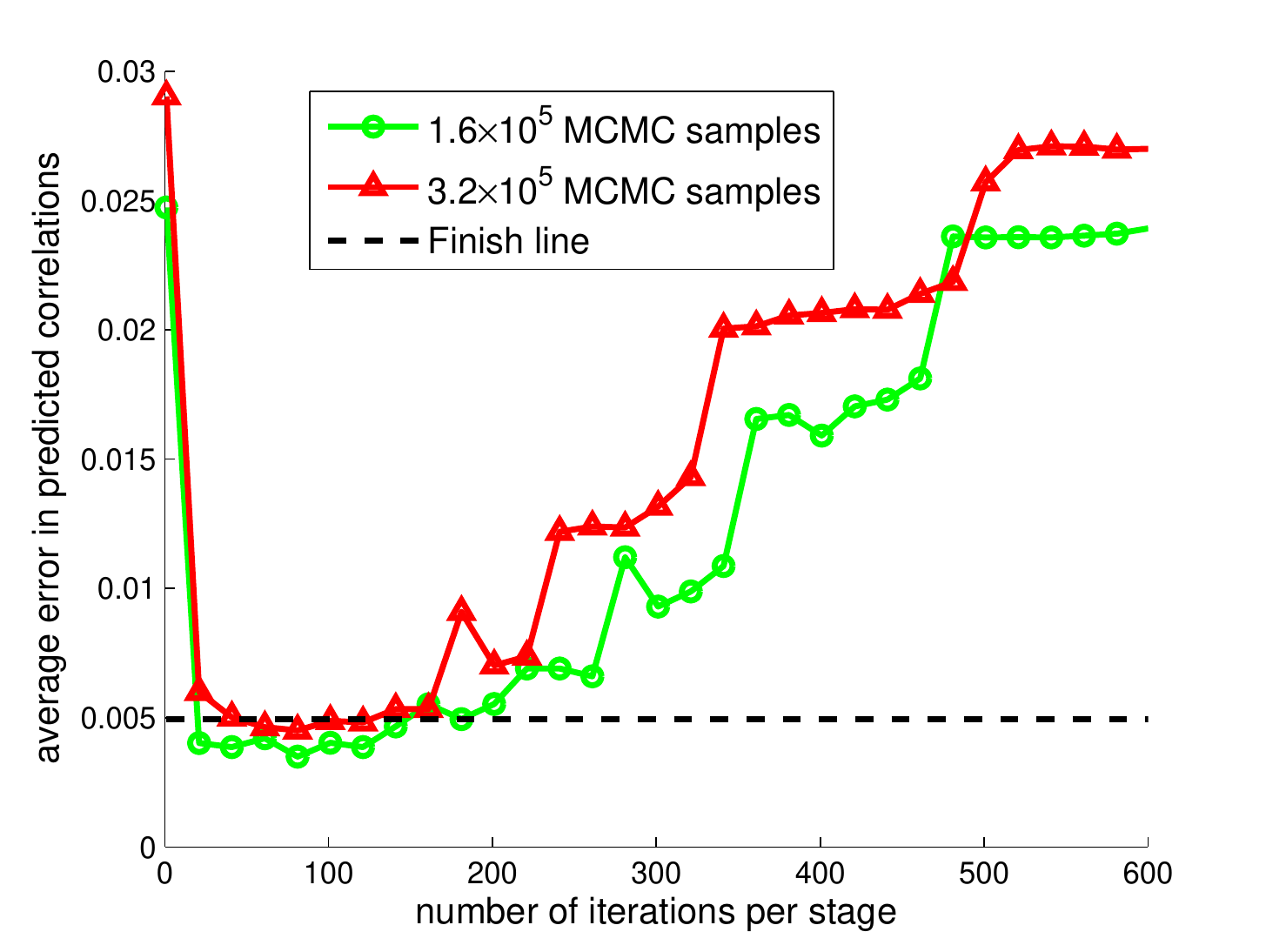}
\hfill
\includegraphics[width=0.45\linewidth]{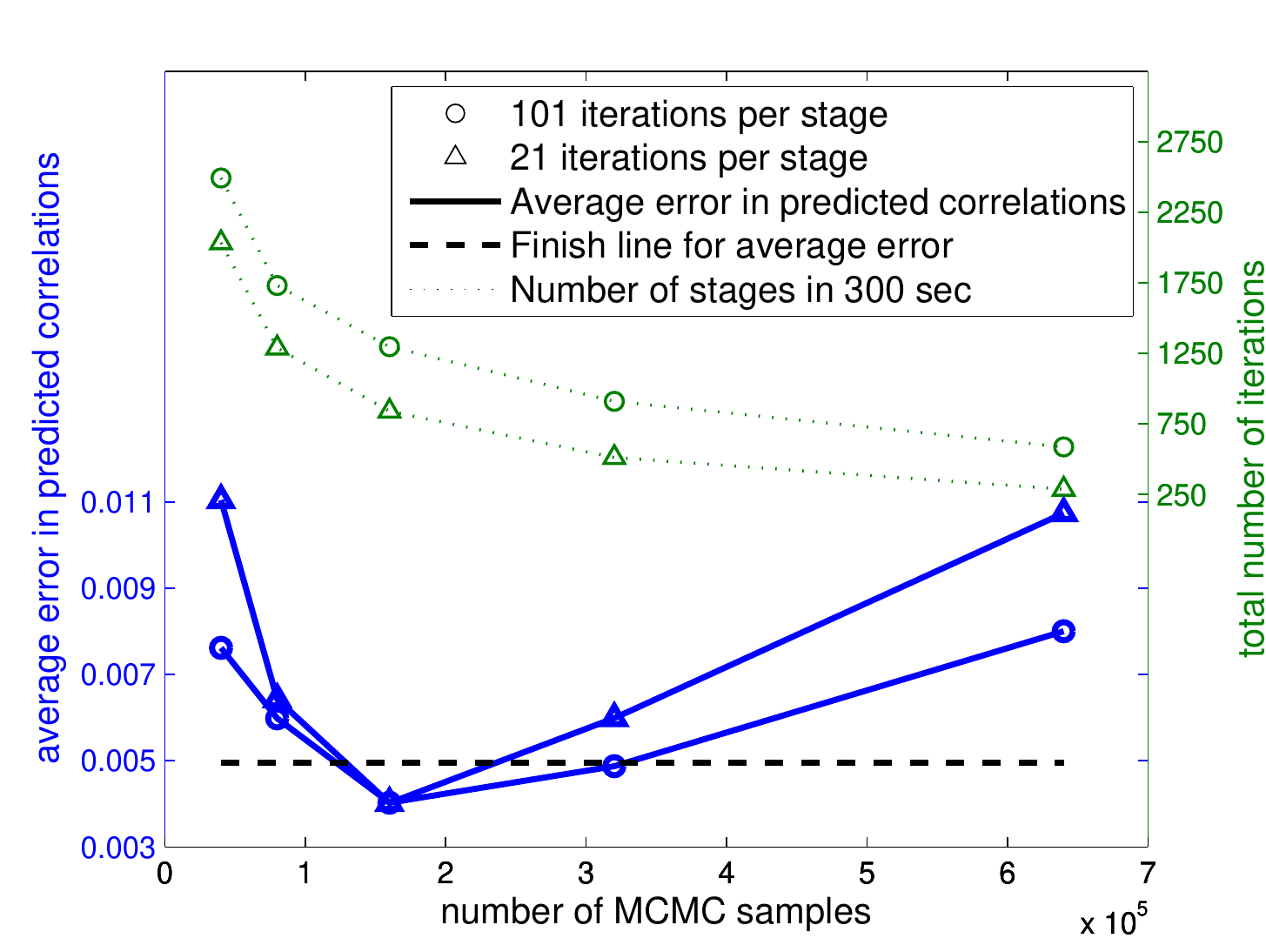}
\end{center}
\caption{\label{fig:c2_compare}
(Left) Performance of our algorithm on the empirical data according
to the absolute correlation difference metric, as a function of the
number of iterations of the optimization algorithm and the number of
Monte Carlo samples. The algorithm was terminated after the running time
exceeded 300 seconds. Choosing too many or too few iterations per
stage decreases efficiency (due to overlearning or inefficient use
of samples, respectively). (Right) Performance of our algorithm on
the empirical data according to the absolute correlation difference
metric, as a function of the number of examples generated by the
Monte Carlo sampler. The algorithm was terminated after the running time
exceeded 300 seconds. Also displayed is the total number of iterations
completed by the time cutoff. Fewer iterations per sample allow the
generation of more samples, which is a computationally expensive
process; therefore, fewer iterations result in fewer total
optimization stages. Choosing too many or too few Monte Carlo samples per
stage decreases efficiency, either because the algorithm samples for
too long and terminates before enough learning stages can be
completed, or because the sampling is too sparse and expectations
are poorly estimated.}
\end{figure*}

\fig{fig:loss_compare} illustrates the performance of the two
algorithms on the synthetic dataset, measured by $\Delta\LL$.
for simplicity, we have held the number of Monte Carlo samples at each stage, $M =3.2\times 10^5$, fixed.
Choosing $T=1$ iteration per stage corresponds to a naive use of Monte Carlo, with a new simulation for each new setting of the parameters, and we see that this approach converges very slowly, as found in Ref \cite{tkacik+al_06}.   Simply increasing this number to $T=20$ produces at least an order of magnitude speed up in convergence.

\subsection{Real data}

Here we consider the problem of constructing the maximum entropy distribution consistent with the correlations observed in real data.  Our example is from Refs \cite{schneidman+al_06,tkacik+al_06}, the responses of $N=40$ retinal neurons.  As noted at the outset, if we divide time into small slices of duration $\Delta\tau$ then each cell either does or does not generate an action potential, corresponding to a binary or Ising variable. The experiments of Ref \cite{schneidman+al_06}, analyzed with $\Delta\tau=0.02\,{\rm s}$, provide $m=189,950$ samples of the state $\vsigma$.  We note that spikes are rare, and pairwise correlations are weak, as discussed at length in Ref \cite{schneidman+al_06}.

As we try to find the parameters $\vtheta$ that describe these data, we need a metric to monitor the progress of our algorithm; of course we don't have access to the true distribution.  Since we are searching in the space of Gibbs distributions which automatically have correct functional form to maximize the entropy, we check  the quality of agreement between the predicted and observed expectation values. It is straightforward to insure that the model reproduces the one--body averages $\langle \sigma_{\rm i}\rangle$ almost exactly.  Thus we focus on the (connected) two--body averages 
$C_{\rm ij} = \langle \sigma_{\rm i}\sigma_{\rm j}\rangle - \langle \sigma_{\rm i}\rangle\langle \sigma_{\rm j}\rangle$.  We compute these averages via Monte Carlo in the distribution $Q_{\vtheta}({\vsigma})$, and then form the absolute difference between computed and observed values of $C_{\rm ij}$, and finally average over all pairs $\rm ij$.  The resulting average error $\Delta C$ measures how close we come to solving the maximum entropy problem.  Note that since the observed values of $C_{\rm ij}$ are based on a finite set of samples, we don't expect that they are exact, and hence we shouldn't identify convergence with $\Delta C \rightarrow 0$.  Instead we divide the real data in half and measure $\Delta C$ between these halves, and use this as the `finish line' for our algorithms.

In Fig \ref{fig:c2_compare} we see the behavior of $\Delta C$ as a function of the number of iterations per stage ($T$), where we terminate the algorithm after 300 seconds of running time on our local cluster \cite{fafner}.  
As expected intuitively, both small $T$ and large $T$ perform badly, but there is a wide range $T\sim 30-200$ in which the algorithm reaches the finish line within the alloted time.  This performance also depends on $M$, the number of Monte Carlo samples per stage, so that again there is a range of $M\sim 1-3\times 10^5$ that seems to work well.  In effect, when we constrain the total run time of the algorithm there is a best way of apportioning this time between stages (in which we run new Monte Carlo simulations) and iterations (in which we adjust parameters using estimates based on a fixed set of Monte Carlo samples).  By working within a factor of two of this optimum, we achieve substantial speed up of the algorithm, or an improvement in convergence at fixed run time.

\section{Conclusion}

Our central conclusion is that recycling of Monte Carlo samples, as in the histogram Monte Carlo method \cite{ferrenberg+swendsen_88}, provides a substantial speedup in the solution of the inverse Ising problem. In detail, some degree of recycling speeds up the computations, while of course too much recycling renders the underlying approximations invalid, so that there is some optimal amount of recycling; fortunately it seems from Fig \ref{fig:c2_compare} that this optimum is quite broad.  We expect that these basic results will be true more generally for problems in which we have to learn the parameters of probabilistic models to provide the best match to a large body of data.  In the specific context of Ising models for networks of real neurons, the experimental state of the art is now providing data on populations of neurons which are sufficiently large that these issues of algorithmic efficiency become crucial.

\begin{acknowledgments}
We thank MJ Berry II, E Schneidman and R Segev for helpful discussions and for their contributions to the work which led to the formulation of the problems addressed here.  This work was supported in part by NIH Grant P50 GM071508, and  by NSF Grants IIS--0613435 and PHY--0650617.
\end{acknowledgments}


\begin{thebibliography}{99}
\bibitem{reweighting}
In early stages of our work we encountered some difficulties with convergence of the coordinate descent algorithm. One possible source of the problem is that if, for example, the Monte Carlo sample includes no spikes from a particular neuron, then trying to match the observed expectation values can become impossible. 
To ease this difficulty we sometimes bias expectation values over the Monte Carlo samples with a small contribution from expectation values taken from the real data.  Further studies would be required to determine if this is essential once we have the optimal values for the parameters $M$ and $T$.
\end{thebibliography}
\end{document}